\def\ltwid{\mathrel{\raise.3ex\hbox{$<$\kern-.75em\lower1ex\hbox{$\sim$}}}}
\documentstyle[multicol,aps,epsf]{revtex}
\begin{document}
\title{Two-Scale Annihilation}
\author{E. Ben-Naim$\dag$, S.~Redner$\ddag$, and P.~L.~Krapivsky$^*$}
\address{$\dag$The James Franck Institute, The University of Chicago, 
Chicago, IL 60637}
\address{$\ddag$Center for Polymer Studies and Department of Physics, 
Boston University, Boston, MA 02215}
\address{$^*$Courant Institute of Mathematical Sciences, 
New York University, New York, NY 10012-1185}
\maketitle
\begin{abstract}
The kinetics of single-species annihilation, $A+A\to 0$, is
investigated in which each particle has a fixed velocity which may be
either $\pm v$ with equal probability, and a finite diffusivity.  In
one dimension, the interplay between convection and diffusion leads to
a decay of the density which is proportional to $t^{-3/4}$.  At long
times, the reactants organize into domains of right- and left-moving
particles, with the typical distance between particles in a single
domain growing as $t^{3/4}$, and the distance between domains growing
as $t$.  The probability that an arbitrary particle reacts with its
$n^{\rm th}$ neighbor is found to decay as $n^{-5/2}$ for
same-velocity pairs and as $n^{-7/4}$ for $+-$ pairs.  These kinetic
and spatial exponents and their interrelations are obtained by scaling
arguments.  Our predictions are in excellent agreement with numerical
simulations.
\end{abstract}

\begin{multicols}{2}

Single-species diffusion-controlled annihilation, $A+A\to 0$, exhibits
classical mean-field kinetics when the spatial dimension $d>2$, in which
the concentration $c(t)$ decays as $t^{-1}$, and nonclassical
dimension-dependent kinetics for $d\leq 2$ with a slower concentration
decay, $c(t)\propto t^{-d/2}$ [1-7].  In one dimension, the geometric
restriction to nearest-neighbor interactions leads to relatively large
departure from the mean-field kinetics, as well as a spatial
organization of reactants.  In this well-studied case, it is known that
$c(t)$ asymptotically decays as $(Dt)^{-1/2}$, independent of the
initial concentration.  The complementary situation of single-species
annihilation where the reactants move ballistically has recently begun
to receive attention [8-12].  Perhaps the simplest example is the
deterministic $\pm$ annihilation process, where each particle moves at a
constant velocity which may be either $+v$ or $-v$ [8,9].  When the
densities of the $+v$ and $-v$ particles are equal, $c(t)$ decays as
$(c_0/vt)^{1/2}$.

In this letter, we consider single species annihilation when the
particle transport is a superposition of convection {\it and} diffusion
--- we term this system the stochastic $\pm$ annihilation process
(Fig.~1).  Although the concentration decays as $t^{-1/2}$ when only one
of the transport mechanisms --- either convection or diffusion --- is
operative, the combined transport process leads to a faster
concentration decay of $t^{-3/4}$.  Our goal is to understand this
unusual decay law and its attendant consequences on the spatial
distribution of reactants.  While there has been fragmentary mention of
some aspects of this system [7,10], here we give primarily new results
and a self-contained account of the basic phenomena.

To set the stage for our approaches and results in the stochastic $\pm$
annihilation process, it is first helpful to provide a simple derivation
for the decay of $c(t)$ in the deterministic $\pm$ process.  Let us
consider a system where particles are placed with concentration $c_0$ in
a box of size $L$, and denote by $c(L,t)$ the time dependent
concentration.  Initially, there are $N=c_0 L$ particles, and the
difference between the number of right and left moving particles is of
the order of $\Delta N=|N_+-N_-|\sim \sqrt{N}$.  Eventually, all
particles who belong to the minority-velocity species are annihilated
and thus $c(L,t=\infty)\sim \Delta N/L\sim (c_0/L)^{1/2}$.  We assume a
scaling from for the concentration, $c(L,t)\sim (c_0/L)^{1/2}f(z)$ with
$z=L/vt$.  According to the above argument, $f(z)\to {\rm const.}$ in
the $z\to 0$ limit.  Conversely, in the short time limit, $z\to\infty$, the
concentration cannot depend on the box size, so that $f(z)$ must be
proportional to $z^{1/2}$.  Thus we find
\begin{equation}
c(t)\sim \left({c_0\over vt}\right)^{1/2}.  
\end{equation}
As a consequence, the system organizes into right- and left-moving
domains whose size is of the order of $vt$.

In the diffusive case, either one particle or no particles survive the
annihilation process in a finite box, depending on the parity of the
initial number of particles.  Following the above line of reasoning, we
may write the scaling ansatz $c(L,t)\sim L^{-1}f(z)$ with
$z=L/\sqrt{Dt}$.  Here the relevant time dependent length scale is
$\sqrt{Dt}$.  In the limit $z\to 0$, the concentration is independent of
$L$, thereby implying $f(z)\sim z$.  Therefore the time dependent
concentration is given by
\begin{equation}
c(t)\sim \left(1\over Dt\right)^{1/2}.  
\end{equation}

The crucial new feature in the stochastic $\pm$ annihilation process
is that particles with the same velocity can mutually annihilate
because of their interaction which is driven by diffusion (Fig.~1).  A
useful way to determine the decay in this process is to consider
separately the role of convection and diffusion on the kinetics.
Because of the convection, particles organize into right-moving and
left-moving domains as outlined above.  Inside each domain, however,
diffusive annihilation between same-velocity particles takes place.
We assume that the diffusive annihilation mechanism leads to an
effective time dependent ``initial'' concentration, $c_0(t)\sim
(Dt)^{-1/2}$, which plays the role of $c_0$ in Eq.~(1).  Thus we obtain
\begin{equation}
c(t)\sim \left(1\over{D\,v^2\,t^3}\right)^{1/4}.
\end{equation}
Intriguingly, the concentration in the stochastic $\pm$ annihilation
process is predicted to decay as $t^{-3/4}$ even though $c(t)$ decays as
$t^{-1/2}$ if either diffusion only or convection only is the transport
mechanism.

\begin{figure}\vspace{-.2in}
\centerline{\epsfxsize=9cm \epsfbox{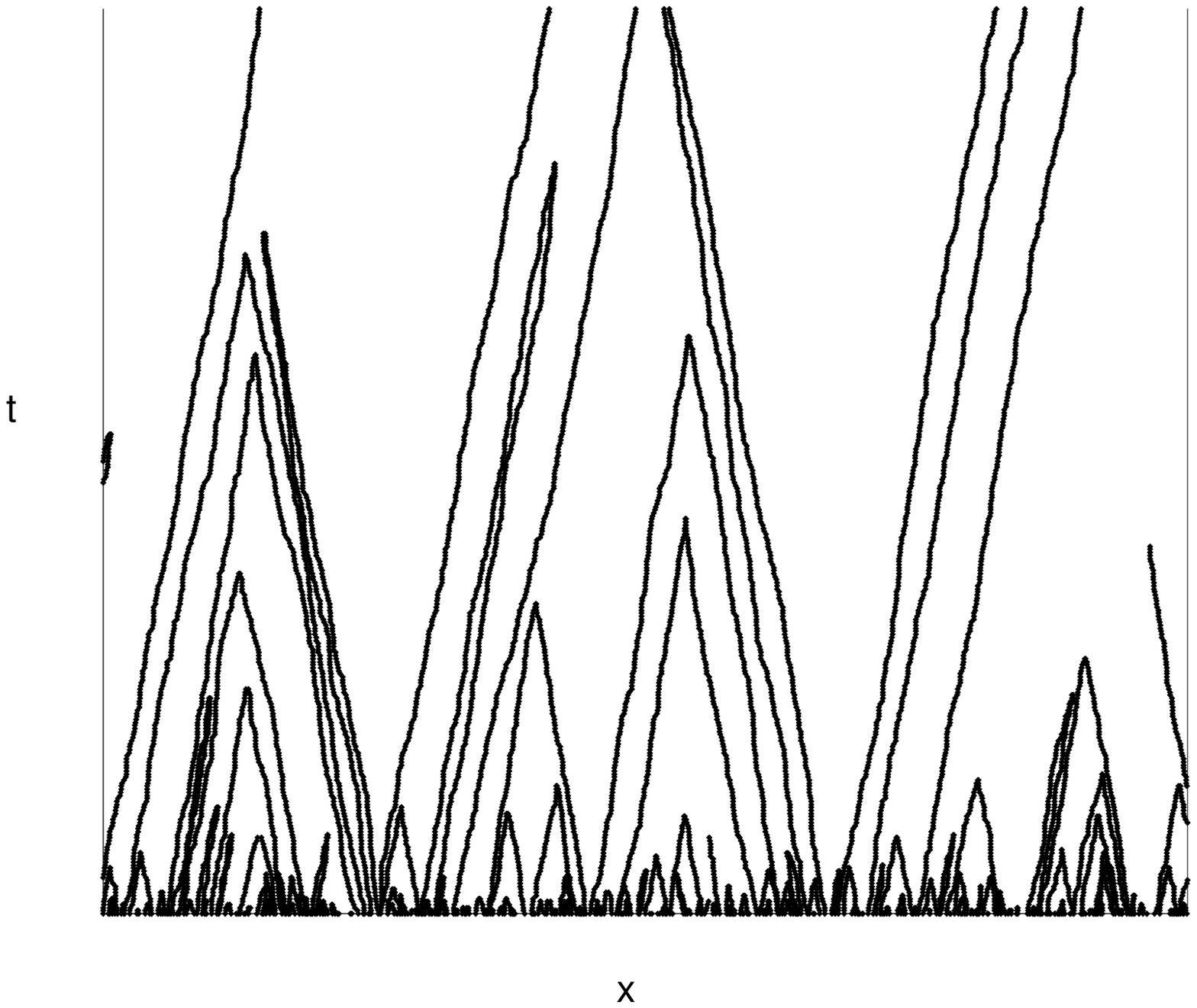}}
\vspace{-.2in}
{\noindent\small Fig. 1 Space-time evolution of particles in the
stochastic $\pm$ annihilation process.}
\end{figure}

An alternative method to determine the decay law, which provides
additional insight into the relative effects of diffusion and
convection, is dimensional analysis.  If the particle diffusion
coefficient is $D$, then the stochastic $\pm$ process is fully
characterized by the initial concentration $c_0$, the velocity $v$, and
$D$.  From these parameters, the only variable combinations with the
dimensions of concentration are $c_0$, ${1/{vt}}$, and ${1/\sqrt{Dt}}$.
On physical grounds, we anticipate that these three concentration scales
should enter multiplicatively so that the time-dependent concentration
can be expressed in a conventional scaling form.  Accordingly, we write
the time dependent concentration in the form
\begin{equation} 
c(t)\sim (c_0)^\rho
\left({1\over{vt}}\right)^\sigma
\left({1\over\sqrt{Dt}}\right)^{1-\rho-\sigma}, 
\end{equation}
in which the dimension of the right-hand side is manifestly a
concentration.  The exponents $\rho$ and $\sigma$ can be now determined
by requiring that the above expression for $c(t)$ matches with: (a) the
diffusion-controlled behavior $c(t)\to (Dt)^{-1/2}$ for $t<\tau_v\simeq
D/v^2$, which is the characteristic time below which the drift can be
ignored for a particle which undergoes biased diffusion; and (b) the
ballistically-controlled behavior $c(t)\to (c_0/vt)^{1/2}$ when
$t<\tau_D\simeq 1/(Dc_0^2)$, which is the time for adjacent particles to
meet by diffusion.  Thus by matching Eq.~(4) with $(Dt)^{-1/2}$ at
$\tau_v$, one obtains $\rho=0$, and then matching Eq.~(4) with
$(c_0/vt)^{1/2}$ at $\tau_D$ gives $\sigma=1/2$.  This then reproduces
Eq.~(3).

To test this decay law, we performed Monte Carlo simulations using the
following realization of the reaction process.  Initially all sites are
occupied with either a $+$ or a $-$ particle with equal probabilities. A
simulation step consists of picking a particle at random and moving it a
single lattice site in the direction of its velocity. If the target site
is occupied, then both particles are removed from the system.  Time is
updated by the inverse of the number of particles.  The simulation was
carried up to $10^5$ time steps on a periodic chain of $10^6$ sites
and an average over $10^3$ realizations was performed.  The data for
$c(t)$ is strikingly linear over a substantial time range on a double
logarithmic scale (Fig.~2).  The local two-point slopes of the data in
the time range $10^2\ltwid t\ltwid 5\times 10^4$ give an exponent value of
$0.745$.  We interpret the constancy of this data as evidence that the
actual value of the exponent is $3/4$.  It is worth noting that a
Pad\'e analysis of the exact short-time power series gives an estimate
for the decay exponent of approximately 0.72 [13].  This provides a
rough estimate for the magnitude of the variation of the effective
exponent between the early time and asymptotic regimes.

\begin{figure}\vspace{-.2in}
\centerline{\epsfxsize=9cm \epsfbox{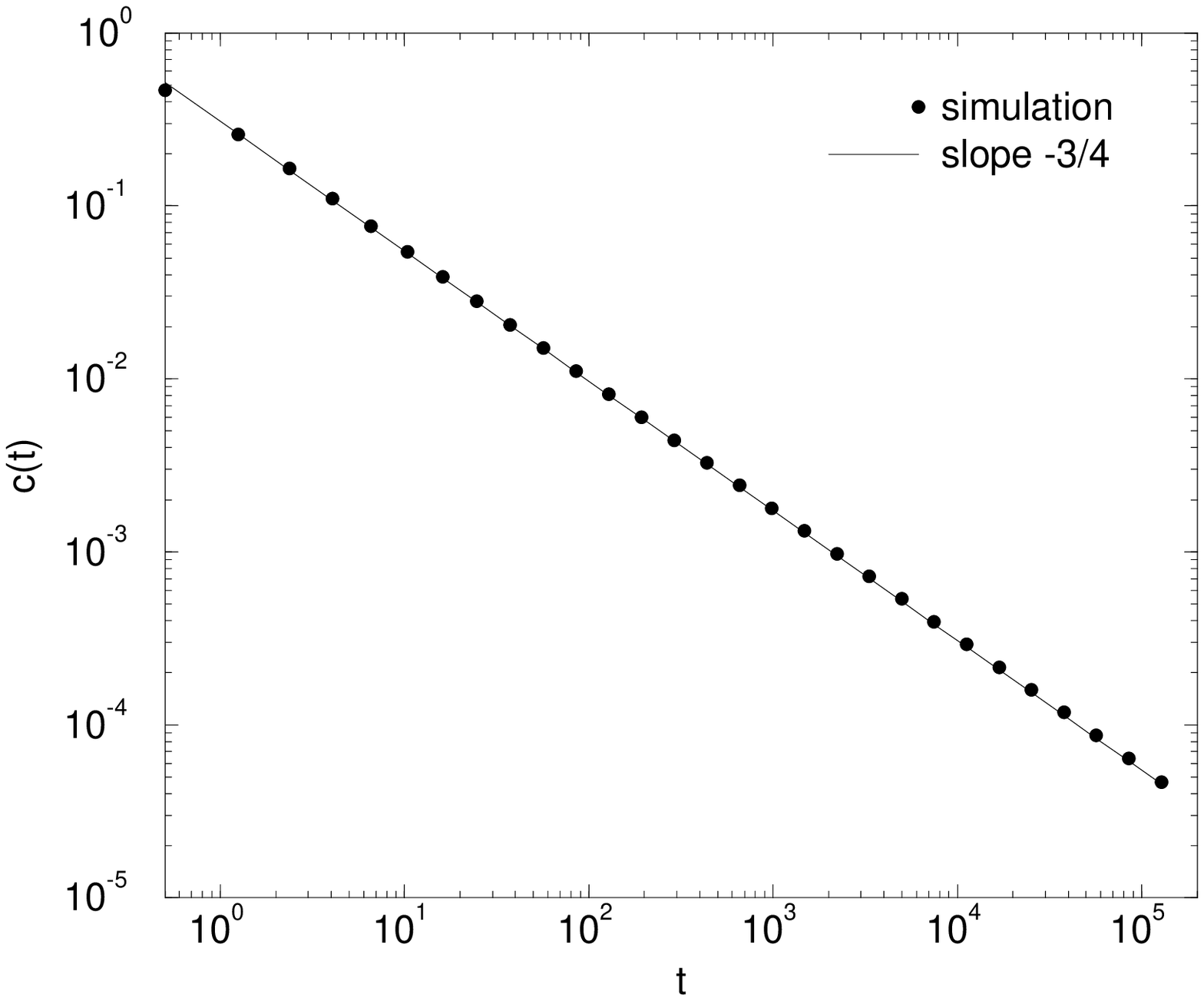}}
\vspace{-.2in}
{\noindent\small Fig. 2 Simulation data for the concentration ($\bullet$) 
versus time on a double logarithmic scale.  A line of slope $-3/4$ is
shown for reference.}
\end{figure}
\begin{figure}
\centerline{\epsfxsize=7.5cm \epsfbox{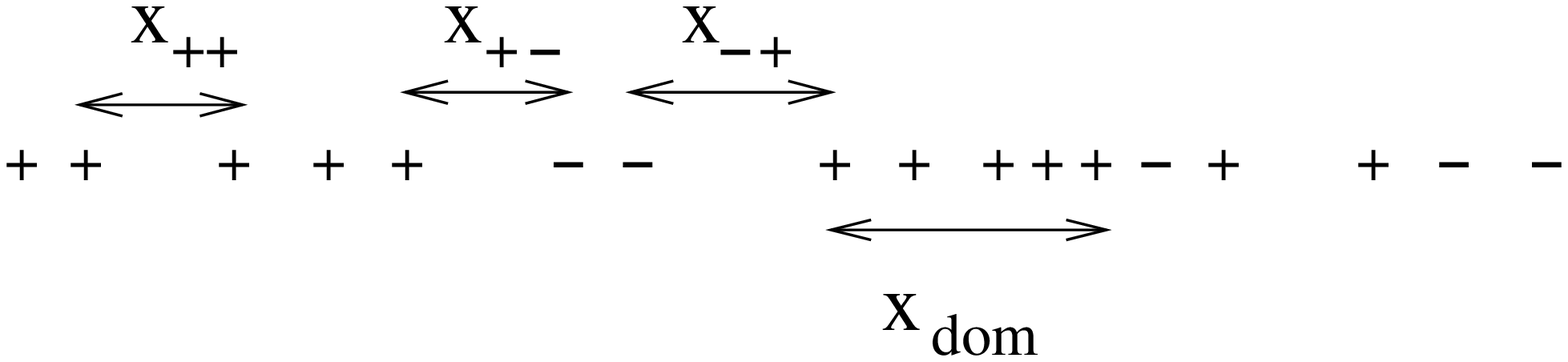}}
\vspace{-.1in}
{\noindent\small Fig. 3 Definition of the basic distance scales that
characterize the spatial organization in the stochastic $\pm$
annihilation process.}
\end{figure}

Having established the decay exponent numerically, it is of interest to
consider the consequences of this unusual decay law on the spatial
distribution of reactants.  In particular, since $c(t)$ decays as
$t^{-3/4}$, one might expect that the average separation between
nearest-neighbor particles grows as $t^{3/4}$.  However, if there
remains any vestige of the domain organization that is associated with
the deterministic $\pm$ process, then more than one length scale may be
needed to characterize this spatial distribution.  Such multiscale
behavior has been observed previously in diffusive two-species
annihilation [14] and the associated consequences lead to new insights
about the system.  To investigate possible multiscale behavior in the
stochastic $\pm$ annihilation process, we introduce the following
distance scales (Fig.~3):
\begin{eqnarray}
\langle{x_{++}(t)}\rangle&\sim t^{\nu_{++}},\quad
\langle{x_{+-}(t)}\rangle&\sim t^{\nu_{+-}},\nonumber\\
\langle{x_{-+}(t)}\rangle&\sim t^{\nu_{-+}},\quad
\langle{x_{\rm dom}(t)}&\rangle\sim t^{\nu_{\rm dom}},
\end{eqnarray} 
which are defined to be, respectively, the average distance between
neighboring same-velocity pairs, $+-$ pairs, $-+$ pairs, and the average
length of a domain of same velocity particles.

\begin{figure}\vspace{-.2in}
\centerline{\epsfxsize=9cm \epsfbox{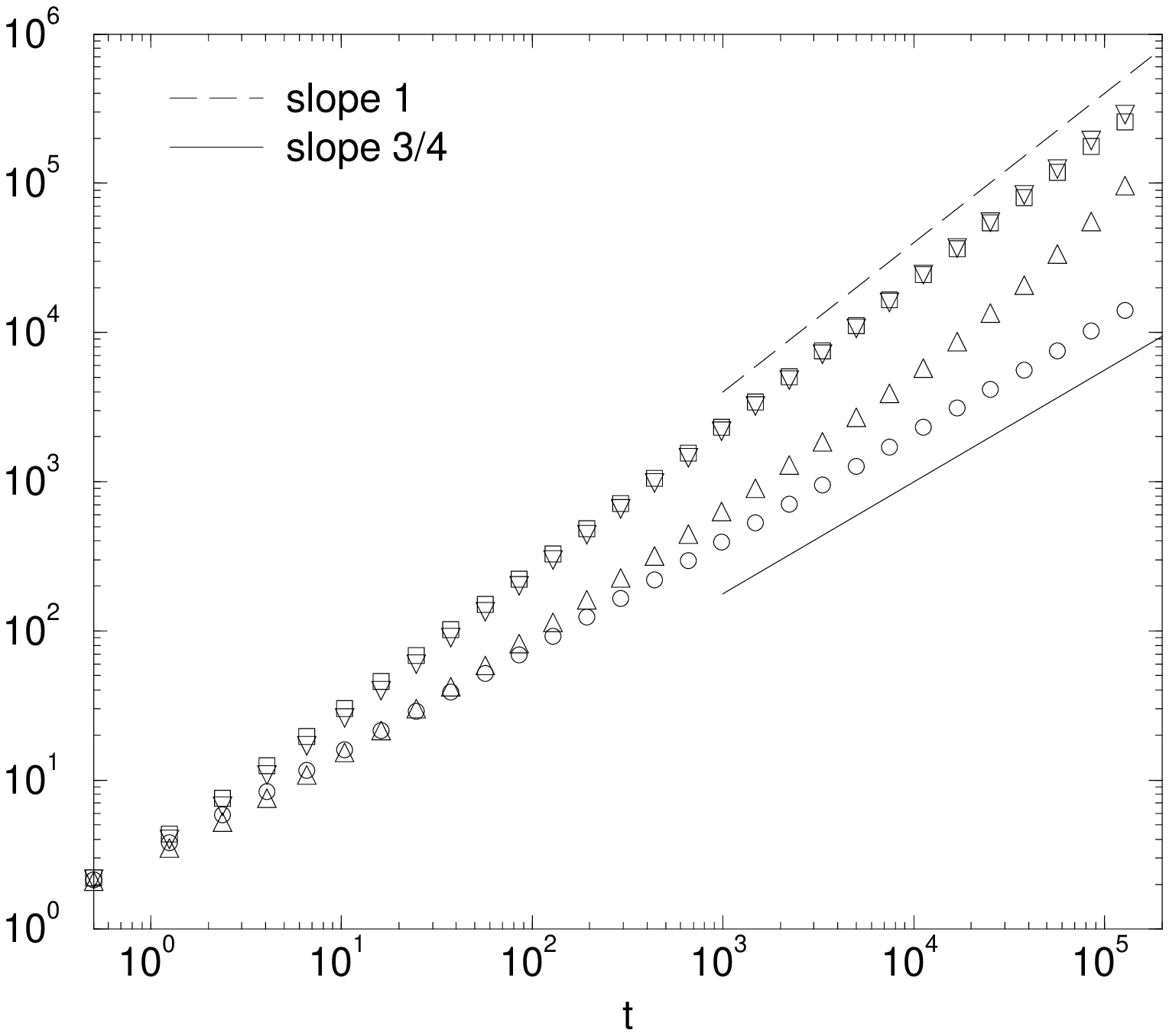}}
\vspace{-.2in}
{\noindent\small Fig. 4 Simulation data for the basic interparticle distances
$\langle{x_{++}(t)}\rangle\sim t^{\nu_{++}}$ ($\circ$),
$\langle{x_{+-}(t)}\rangle\sim t^{\nu_{+-}}$ ($\bigtriangleup$),
$\langle{x_{-+}(t)}\rangle\sim t^{\nu_{-+}}$ ($\Box$), and
$\langle{x_{\rm dom}(t)}\rangle\sim t^{\nu_{\rm dom}}$
($\bigtriangledown$).  Lines of slopes $3/4$ and $1$ are also shown
for reference.}
\end{figure}

Our Monte Carlo data for these length scales exhibits considerable
curvature on a double logarithmic scale (Fig.~4).  Thus to estimate
the asymptotic behavior, we studied the systematic variation of the
slopes of linear least-squares fits as the data at the earliest times
are progressively eliminated.  The effective exponents obtained in
this manner vary considerably; for example, for
$\langle{x_{++}(t)}\rangle$, the effective exponent systematically
increases, but at a progressively slower rate, from 0.699 to 0.734.
Together with relatively strong numerical evidence that the
concentration decays as $t^{-3/4}$, we conclude that the actual value
of $\nu_{++}$ is $3/4$.  This accords with the expectation that
$\langle{x_{++}(t)}\rangle$ should scale as $1/c(t)$.  Similar finite
time corrections occur in the exponent estimates for the remaining
length scales defined above.  For these cases, the effective exponent
values are all increasing as short-time data are systematically
deleted and it appears that $\nu_{+-}$, $\nu_{-+}$, and $\nu_{\rm
dom}$ are all very close to 1, asymptotically.  That is, the
corresponding lengths are governed by the ballistic particle motion,
but again with considerable finite time corrections.  The case of
$\langle{x_{+-}(t)}\rangle$ is especially problematic, as the effective
exponent changes from approximately $0.80$ to $0.93$ over the time
range covered by our simulation.  Evidently, more extensive simulation
would be needed to determine the asymptotic exponent values
unambiguously by simulation alone.

A new useful way to characterize the spatial range of bimolecular
reactions is the collision probability, $P(n)$, defined as the
probability that the reaction partner of a given particle is its $n^{\rm
th}$ neighbor.  Eventually, every particle reacts with some collision
partner in one dimension and the distribution of the distances between
partners provides a measure of the reaction ``efficiency''.  In the
deterministic $\pm$ process, for example, this probability can be
obtained analytically [8,9,15].  Let us denote the velocity of the
$n^{\rm th}$ neighbor by $v_n=\pm 1$, and the local velocity sum by
$S_n=\sum_{i=0}^n v_i$.  A right moving particle initially at the origin
reacts with its $(2n+1)^{th}$-neighbor if: (a) $S_l>0$ for
$l=0,1,\ldots,2n$, and (b) $S_{2n+1}=0$.  This quantity is precisely the
same as the first-passage probability for a random walk which starts at
the origin to return to the origin for the first time after $2n$ steps.
Because of this equivalence to an exactly soluble first-passage problem
[16], one has $P(2n)=0$ and $P(2n+1)=2^{-2n-1}{(2n)!/ n!(n+1)!}$.  In
the limit $n\to\infty$, the probability that a given particle collides
with its $n^{\rm th}$-neighbor is given by
\begin{equation}
P(n)\propto n^{-3/2}.
\end{equation}

Motivated by this power law dependence, we assume, in general, that
$P(n)\sim n^{-\gamma}$.  The exponent $\gamma$ can be related to other
fundamental exponents of reaction processes, namely, the concentration
decay exponent $\alpha$, defined by $c(t)\sim t^{-\alpha}$, and the
correlation exponent $\beta$, defined by $\xi(t)\sim t^{\beta}$.  Here
$\xi(t)$ refers to the distance over ``information'' about the reactants
spread.  In a time $t$, only particles within a domain of linear size
$\xi(t)$ are eligible to react and thus the surviving fraction, or
concentration, is
\begin{equation} 
c(t)\sim \int_{\xi(t)} dn\,
P(n)\sim \int_{\xi(t)} dn\, n^{-\gamma}
\sim t^{\beta(1-\gamma)}. 
\end{equation}
Consequently, we find the exponent relation 
\begin{equation}
\gamma=1+\alpha/\beta.
\end{equation}
For the deterministic $\pm$ process, $\alpha=1/2$ and $\beta=1$ [8,9], and
the exact $\gamma=3/2$ of Eq.~(6) is recovered.  As an illustration, consider,
for example, single-species diffusion-limited annihilation.  The decay
and correlation exponents are $\alpha=1/2$ and $\beta=1/2$, leading to
$\gamma=2$ from Eq.~(8).  Preliminary simulations appear to confirm this
result.  Similarly, for two-species annihilation, $\alpha$ is now
equal to 1/4 while $\beta$ remains 1/2 so that $\gamma=3/2$.

Let us now consider the behavior of the collision probability in the
stochastic $\pm$ annihilation process.  In this case, the existence of
two length scales in the system suggests that it is necessary to make a
distinction between reaction events that involve particles of the same
and of different velocities.  We therefore define a ballistic
correlation scale $\xi_{+-}(t)\sim t^{\beta_{+-}}$, with $\beta_{+-}=1$,
which is associated with $+-$ collisions, {\it i.e.}, annihilation events
between opposite velocity particles.  Invoking the scaling relation
Eq.~(8), we thus find $P_{+-}(l)\sim l^{-\gamma_{+-}}$ with
$\gamma_{+-}=7/4$.  Similarly, there is a diffusive length scale
$\xi_{++}(t)\sim t^{\beta_{++}}$, with $\beta_{++}=1/2$, corresponding
to annihilation events between same-velocity particles.  In this case,
Eq.~(8) gives $\gamma_{++}=5/2$.  To summarize, we obtain
\begin{equation}
P(n)\sim P_{+-}(n)\sim n^{-3/4}\quad 
P_{++}(n)=P_{--}(n) \sim n^{-5/2}.
\end{equation}
This behavior is consistent with our Monte Carlo simulation data
(Fig.~5).  Notice that over large distances, annihilation between
opposite velocity particles dominates, as one would naively expect.

\begin{figure}\vspace{-.2in}
\centerline{\epsfxsize=9cm \epsfbox{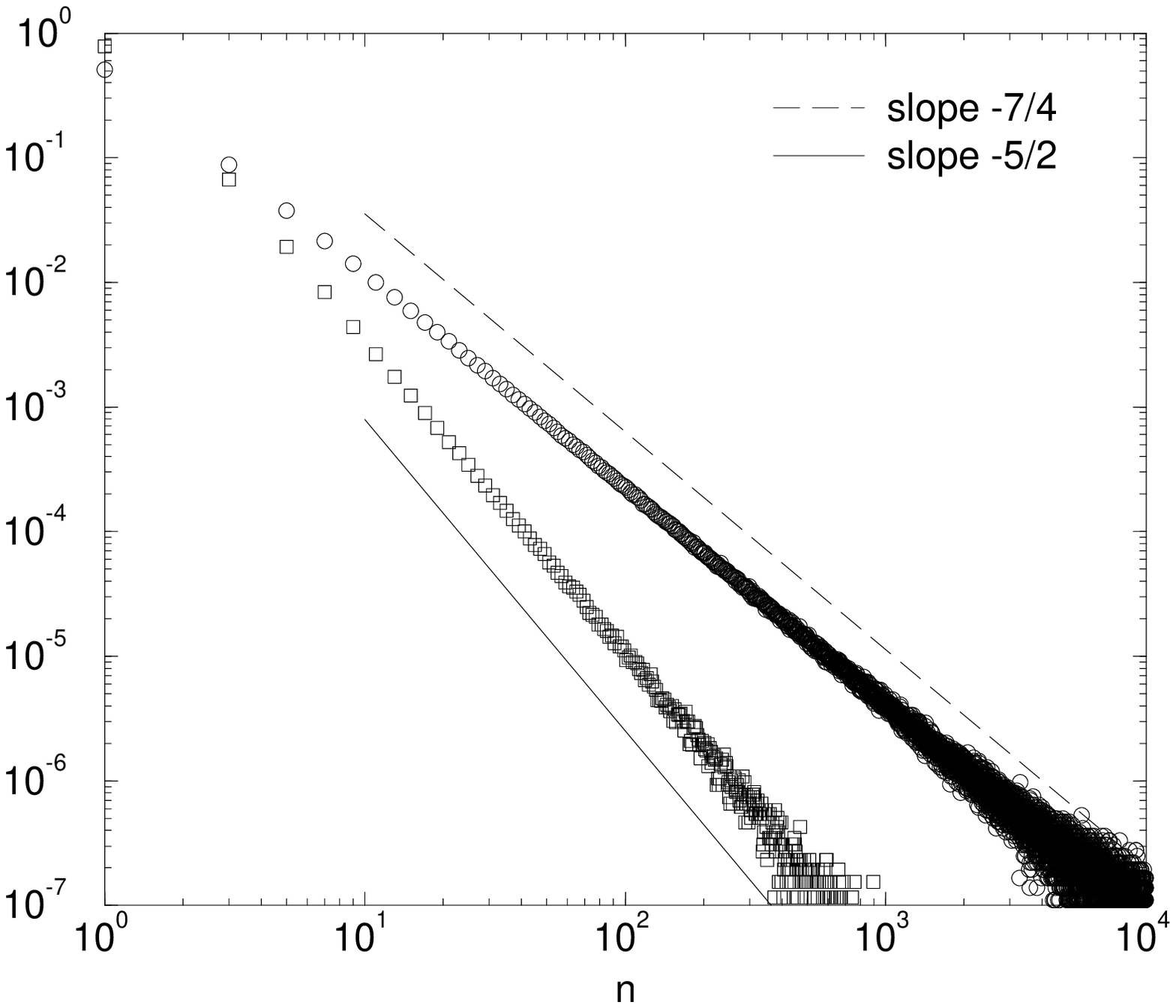}}
\vspace{-.2in}
{\noindent\small Fig. 5 Simulation data for $P_{+-}(n)$ ($\circ$) and
$P_{++}(n)$ ($\Box$) on a double logarithmic scale.  Lines of
slopes $-7/4$ and $-5/2$ are drawn as guides to the eye.}
\end{figure}

Our results can also be generalized to arbitrary spatial dimension
$d>1$.  In this case, it is necessary to ascribe a finite, non-zero
radius $R$ to the particles so that there is a finite collision cross
section for particles to actually meet.  Let us consider the
anisotropic system in which particles undergo isotropic Brownian
motion with diffusivity $D$, and a drift along the $\hat x$ axis only,
with the velocity taking on the value $\pm v \hat x$ with equal
probability.  In the ballistic limit ($D\equiv 0$), the concentration
decays as $\sqrt{c_0/R^{d-1}vt}$ (since the process is
quasi-one-dimensional, the $t^{-1/2}$ decay of the true
one-dimensional system is still obeyed).  In contrast, for
diffusion-controlled annihilation ($v\equiv 0$), the concentration
decays as $(Dt)^{-d/2}$ for $d<2$, and as $(R^{d-2}Dt)^{-1}$ for $d>2$
(with logarithmic corrections at the critical dimension $d=2$) [7].
Repeating the analysis detailed previously for the one-dimensional
case in the derivation of Eq.~(3) from Eqs.~(1) and (2), we find the
concentration decay
\begin{equation}
c(t)\sim \cases{
(R^{2-2d}\,D^d\,v^2)^{-1/4}\,t^{-(d+2)/4}                        &$d<2$;\cr
(R\,D\,v)^{-1/2}\,t^{-1}\,\left[\ln(Dt/R^2)\right]^{1/2},  &$d=2$;\cr
(R^{2d-3}\,D\,v)^{-1/2}\,t^{-1},                               &$d>2$.\cr}
\end{equation}
The combined diffusion and ballistic transport does not change the
mean-field nature of the annihilation kinetics when $d>2$ and the
classical $t^{-1}$ decay is recovered.  For sufficiently low spatial
dimension, however, the non-classical behavior arises in which the decay
exponent $\alpha=(d+2)/4$.  Thus in low spatial dimensions, the
interplay between convection and diffusion provides more effective
mixing than diffusion or drift alone, and leads to a larger decay
exponent than $\alpha_{\rm diff}=d/2$ and $\alpha_{\rm ball}=1/2$ 
which arise when only one transport mechanism is operative.

In summary, the stochastic $\pm$ single-species annihilation process
exhibits a $t^{-3/4}$ decay of the concentration.  This is faster than
the $t^{-1/2}$ decay that arises when only one of the constituent
transport processes in the stochastic $\pm$ process, either diffusion
and deterministic $\pm$ convection, is present.  A microscopic
understanding of this decay law is lacking, and it seems that a
technique beyond those typically used to solve one-dimensional
reactive systems would be needed for the stochastic $\pm$ annihilation
process.  At long times, the system exhibits a spatial organization in
which diffusion controls the short distance behavior and convection
controls the large distance behavior.  We have also introduced the
concept of the collision probability, $P(n)$, the probability that a
given particle is annihilated by its $n^{\rm th}$-neighbor.  For the
stochastic $\pm$ process, this probability is further discriminated by
annihilation by same-velocity and opposite velocity pairs.  These two
probabilities decay as $P_{++}(n)\sim n^{-5/2}$ and $P_{+-}(n)\sim
n^{-7/4}$, respectively.  It will be interesting to study the
collision probability in other reaction processes such as diffusive
driven single-species annihilation.

\vspace{.1in}
We gratefully acknowledge support from the NSF under awards 92-08527,
MRSEC program DMR-9400379 (EBN), DMR-9219845 and ARO grant
DAAH04-93-G-0021 (SR). PLK was supported in part by a grant from NSF.

\end{multicols} 

\begin{thebibliography}{99}


\bibitem{} M.~Bramson and D.~Griffeath,
      {\sl Z. Wahrsch. verw. Gebiete} {\bf 53}, 183 (1980).

\bibitem{}  D.~C.~Torney and H.~M.~McConnell, 
      {\sl Proc. Roy. Soc. London A} {\bf 387}, 147 (1983).

\bibitem{} Z.~Racz, {\sl Phys. Rev. Lett.} {\bf 55}, 1707 (1985).

\bibitem{}  A.~A.~Lushnikov, 
      {\sl Sov. Phys. JETP} {\bf 64}, 811 (1986).

\bibitem{} J.~L.~Spouge, 
      {\sl Phys. Rev. Lett.} {\bf 60}, 871 (1988).  

\bibitem{} D.~ben-Avraham, M.~A.~Burschka and C.~R.~Doering, 
      {\sl J. Stat. Phys.} {\bf 60}, 695 (1990).

\bibitem{} For a recent review of diffusion-controlled annihilation, see  
      S.~Redner, 
      in {\it Nonequilibrium Statistical Mechanics in One Dimension}, 
      ed. V.~Privman (Cambridge University Press, Cambridge, 1996).

\bibitem{} Y.~Elskens and H.~L.~Frisch, 
      {\sl Phys. Rev. A} {\bf 31}, 3812 (1985).  

\bibitem{} J.~Krug and H.~Spohn, 
      {\sl Phys. Rev. A} {\bf 38}, 4271 (1988).

\bibitem{} E.~Ben-Naim, S.~Redner, and F.~Leyvraz, 
      {\sl Phys. Rev. Lett.} {\bf 70}, 1890 (1993).

\bibitem{} J.~Piasecki, 
      {\sl Phys. Rev. E} {\bf 51}, 5535 (1995).  

\bibitem{} M.~Droz, P.-A.~Rey, L.~Frachebourg, and J.~Piasecki, 
      {\sl Phys. Rev. E} {\bf 51}, 5541 (1995).

\bibitem{} E.~Ben-Naim and J.~Zhuo, 
      {\sl Phys. Rev. E} {\bf 48}, 2603 (1993).

\bibitem{} F.~Leyvraz and S.~Redner,  
      {\sl Phys. Rev. E} {\bf 66}, 2168 (1991).

\bibitem{} P.~L.~Krapivsky, S.~Redner, and F.~Leyvraz, 
      {\sl Phys. Rev. E} {\bf 51}, 3977 (1995).

\bibitem{} W.~Feller, 
      {\sl An Introduction to Probability Theory and it Applications}, 
      Vol I (Wiley, New York, 1968).  
\end{thebibliography}
\end{document}